\documentclass[aps,twocolumn,groupedaddress,showpacs]{revtex4}

\begin{document}
\bibliographystyle{apsrev}


\title{Quantum Order: a Quantum Entanglement of Many Particles}


\author{Xiao-Gang Wen}
\homepage{http://dao.mit.edu/~wen}
\affiliation{Department of Physics, Massachusetts Institute of Technology,
Cambridge, Massachusetts 02139}


\date{Sept. 2001}

\begin{abstract}
It is pointed out that quantum states, in general, contain a new kind of orders
that cannot be characterized by symmetry.  A concept of quantum order is
introduced to describe such orders. As two concrete examples, we discussed
quantum orders in free fermion systems and in 2D spin-1/2 systems.  We
generalize the Landau's theory for the classical orders and introduce
projective symmetry group and Fermi surface topology to characterize different
quantum orders.
\end{abstract}

\maketitle


{\bf Introduction}:
Symmetry breaking and the associated order parameter have been playing a key
role in our understanding of phases  and phase transitions.\cite{LL58,GL5064}
However, recent study of quantum Hall (QH) liquids reveal that QH liquids
contain a new kind of order -- topological order -- 
which can not be characterized by broken symmetry
and the associated order parameter.\cite{Wtop,WNtop} 
Thus Landau's symmetry breaking theory for
phase and phase transition does not apply to QH liquids and a new theory    
was developed to describe the topological orders in QH
liquids.\cite{Wtoprev}

The reason that the Landau's symmetry breaking theory does not apply to QH
liquids is because the Landau's theory was developed for classical statistical
systems which are described by {\em positive} probability
distribution functions of infinite variables. The QH liquids are described
by their ground state wave functions which are {\em complex} functions of
infinite variables. Thus it is not surprising that QH liquids contain addition
structures (or a new kind of orders) that cannot be described by broken
symmetries and the Landau's theory. From this point of view, we conclude that
{\em any} quantum states may contain new kind of orders that are beyond broken
symmetry characterization. Such kind of orders will be called quantum
order.\cite{Wqo} 

To visualize the distinction between the classical order and the quantum order,
we may view the classical world described by positive probabilities as a
``black and white'' world, while the quantum world described by complex wave
functions as a ``colorful'' world.  The Landau's theory based on symmetry
principle and order parameters is color blind which can only describe
classical orders.  We need to use new theories, such as the theory of
topological/quantum orders, to characterize the rich ``color'' in quantum world.
We can also view quantum order as a description of the pattern 
of the quantum entanglement in a many-body ground state.  A special collective
excitation above a quantum ordered state - gauge fluctuations - can be viewed
as the fluctuations of quantum entanglement.  In contract, the classical order
in a crystal just describes a static positional pattern, which has no
non-trivial quantum entanglement.

{\bf Quantum phase transitions and quantum orders}:
Classical orders can be studied through classical phase transitions.
Classical phase transitions are marked by singularities in the free energy
density $f$. The free energy density can be calculated through the partition
function:
\begin{align}
\label{ClZ}
 f =& -\frac{T\ln Z}{V_{space}},\ \ \ \
 Z = \int D\phi e^{-\bt \int dx h(\phi)}
\end{align} 
where $h(\phi)$ is the energy density of the classical system and $V_{space}$
is the volume of space.

Similarly, to study quantum orders, we need to study quantum phase transition
at zero temperature $T=0$. Here the energy density of the ground state
play the role of free energy density. A singularity in the ground state energy
density marks a quantum transition. The similarity between the ground state
energy density and the free energy density can be seen clearly in the
following expression of the energy density of the ground state:
\begin{align}
\label{QuZ}
 \rho_E =& i\frac{\ln Z}{V_{spacetime}},\ \ \ \
 Z = \int D\phi e^{i\int dxdt \cL(\phi)}
\end{align} 
where $\cL(\phi)$ is the Lagrangian density of the quantum system and
$V_{spacetime}$ is the volume of space-time.  We also note that the free
energy density becomes the ground state energy density at $T=0$.  Comparing
\Eq{ClZ} and \Eq{QuZ}, we see that a classical system is described by a path
integral of a positive functional, while a quantum system is described by a
path integral of a complex functional.  This is the real reason why
the classical and quantum orders are different.  According to the
point of view of quantum order, a quantum phase transition, marked by a
singularity of the path integral of a complex functional, in general, cannot
be characterized by a change of symmetry and the associated order parameter.
Thus, in general, we cannot use the broken symmetry and the Ginzburg-Landau
theory to describe a continuous quantum transition.

Although the above discussion is limited to zero temperature, the path
integrals of some quantum systems can be complex even at finite temperatures.
Thus the above result also apply to quantum systems at finite temperatures. It
is possible that a continuous phase transition of a quantum system also does
not involve any change of symmetry even at finite temperatures. 

{\bf Quantum orders and quantum transitions in free fermion systems}:
Let us consider free fermion system with only the translation symmetry and the
$U(1)$ symmetry from the fermion number conservation.  
The Hamiltonian has a form
\begin{equation}
 H=\sum_{\<\v i\v j\>} \left(c^\dag_{\v i} t_{\v i\v j}c_{\v j} 
                             + h.c.\right)
\end{equation}
with $t_{\v i\v j}^*=t_{\v j\v i}$.  
The ground state is obtained by filling every negative energy state with one
fermion.  In general, the system contains several pieces of Fermi surfaces.



\begin{figure}
\centerline{
\includegraphics[width=3.2in]{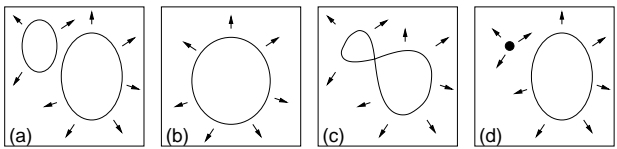}
}
\caption{
Two sets of oriented Fermi surfaces in (a) and (b) 
represent two different quantum orders. 
The two possible transition points between
the two quantum order (a) and (b) are
described by the Fermi surfaces (c) and (d).
}
\label{trans2d}
\end{figure}

To understand the quantum order in the free fermion ground state, we note that
the topology of the Fermi surfaces can changes in two ways as we continuously
changing $t_{ij}$: 
(a) a Fermi surface shrinks to zero (Fig. \ref{trans2d}d) and
(b) two Fermi surfaces join (Fig.  \ref{trans2d}c).
When a Fermi surface is about to disappear in a $D$-dimensional system, the
ground state energy density has a form
\begin{equation}
 \rho_E = \int \frac{d^D\v k}{(2\pi)^D} (\v k\cdot M \cdot \v k -\mu)
\Th( -\v k\cdot M \cdot \v k+\mu) + ...  \nonumber 
\end{equation}
where the $...$ represents non-singular contribution and the symmetric matrix
$M$ is positive (or negative)
definite.  
We find that the ground state energy density has a singularity at $\mu=0$:
$ \rho_E = c \mu^{(2+D)/2} \Th(\mu) + ...$,
where 
$ \Th(x>0) = 1$, $\Th(x<0) = 0$.
When two Fermi surfaces are about to join, the singularity is still
determined by the above equation, but now
$M$ has both negative and positive eigenvalues.
The ground state energy density has a singularity
$ \rho_E = c \mu^{(2+D)/2} \Th(\mu) + ...$ when $D$ is odd and
$ \rho_E = c \mu^{(2+D)/2} \log |\mu| + ...$ when $D$ is even.

We find that the ground state energy density has a singularity at $\mu=0$
which is exactly the same place where the topology of the Fermi surfaces has a
change.\cite{L6030} Thus the topology of the oriented Fermi surface is a
``quantum number'' that characterizes the quantum order in a free fermion
system.  (see Fig.  \ref{trans2d}).  A change in the topology signals a {\em
continuous} quantum phase transition. Lifshitz\cite{L6030}
also studied critical properties of such quantum transitions for $D=3$.

%

{\bf Quantum order in spin liquids}:
Quantum order simply represents the quantum entanglement in ground state.
The ground state wave function of a free fermion system has a simple form of
Slater determinant. Its quantum order can be represented by
the topology of the Fermi surfaces. In this section, we are going to discuss
quantum orders in spin liquids. The ground state wave functions for spin
liquids are more complicated and cannot be written as a Slater determinant.
Thus we need to find a new way to characterize the quantum order in spin
liquids.

The problem here is that the many-body wave function is too complicated to
write down and it is hard to study them without writing them down. 
To gain some insights into the quantum entanglements in spin liquid states,
in the following, we will discuss a concrete representation of the spin wave
function. We will consider a spin-1/2 system on a 2D square lattice. For such a
system, the spin wave function can be written as a bosonic wave function
$\Phi(\v x_1,...,\v x_{N_{up}})$, where $\v x_i$ is the coordinate of the
$i^{th}$ up-spin and $N_{up}$ is the total number of the up-spin. 

Within the $SU(2)$ slave-boson approach,\cite{AZH8845,DFM8826}
instead of writing down the bosonic
wave function $\Phi(\{\v x_i\})$ directly, we regard the boson as a bound
state of two fermions $\psi_1$ and $\psi_2$ (which will be called spinons) and
write $\Phi$ as
\begin{equation}
\label{PhiPsi12}
\Phi(\{\v x_i\}) = \Psi_1(\{\v x_i\}) \Psi_2(\{\v x_i\}) 
\end{equation}
where $\Psi_{1,2}$ are the wave functions of $\psi_{1,2}$ and have a form of
Slater determinant.\cite{J9153,Wpc} Actually, more general spin wave function
can be constructed by introducing a mean-field Hamiltonian
\begin{equation}
\label{Hmean}
 H_{mean}= \sum_{\v i\v j} \psi^\dag_{\v i} u_{\v i\v j} \psi_{\v j}
+ \sum_{\v i} \psi^\dag_{\v i} a_0^l(\v i) \tau^l \psi_{\v i}
\end{equation}
where $\psi^T=(\psi_1,\psi_2)$, $\tau^{1,2,3}$ are the Pauli matrices, and
$u_{\v i\v j}$ are $2\times 2$ complex matrices. The collection $(u_{\v i\v
j}, a_0^l\tau^l)$ is called a mean-field ansatz. For each mean-field ansatz,
we can obtain a mean-field ground state by filling the lowest $2N_{up}$
energy levels of $H_{mean}$ with the spinons $\psi$:
$|\Psi_{mean}^{(u_{\v i\v j}, a_0^l\tau^l)}\> =
 \Psi_{mean}^{(u_{\v i\v j}, a_0^l\tau^l)}(\v y_1,..,\v z_1,..)$,
where $\v y_i$ are the coordinates for $\psi_1$ and $\v z_i$ for $\psi_2$. The
physical spin wave function $\Phi(\{\v x_i\})$ can now be obtained by
performing a projection\cite{G6359} $\v y_i=\v z_i = \v x_i$:
\begin{align}
\label{PhiPsi}
 \Phi^{(u_{\v i\v j}, a_0^l\tau^l)}(\v x_i) =|_{\v y_i=\v z_i = \v x_i}
 \Psi_{mean}^{(u_{\v i\v j}, a_0^l\tau^l)}(\v y_i,\v z_i)
\end{align}
\Eq{PhiPsi} generalizes \Eq{PhiPsi12}.  What we have achieved here is that we
manage to construct a large class of spin wave functions. Those spin wave
functions $\Phi$ are related, through the projection, to the mean-field wave
function $\Psi$ which have a simple form of Slater determinant. The
constructed spin wave function can be labeled by the mean-field ansatz $(u_{\v
i\v j}, a_0^l\tau^l)$.  This allows us to study the quantum
order of a spin wave function by studying the property of a simpler object,
the mean-field ansatz $(u_{\v i\v j}, a_0^l\tau^l)$.

In the study of phases and the related internal orders, the central question is
to identify the universal properties of states. By definition, a universal
property is a property shared by all the states in the same phase.  
For classical systems, the symmetry of a state is a universal property.  We
cannot change one state to another state without a phase transition if the two
states have different symmetries.  Therefore, we can use the symmetry group
(SG) to characterize the internal orders of classical states.

Similarly, to characterize quantum orders, we need to find the universal
properties of spin liquid wave function. Using the above projective
construction, we can simplify the problem by considering the universal
properties of the mean-field ansatz $(u_{\v i\v j}, a_0^l\tau^l)$ instead.
Motivated by the classical systems, here we would like to propose that the
symmetry of the mean-field ansatz $(u_{\v i\v j}, a_0^l\tau^l)$ is a universal
property.  The symmetry group of the ansatz will be called the projective
symmetry group (PSG).  Under our conjecture, the quantum orders in spin
liquids can be characterized by PSG's.  

At the first sight, it appears that the PSG characterization of quantum orders
is identical to the SG characterization of classical order used in
Landau's theory. In fact the PSG of a spin wave function is different from the
SG of that wave function and the two characterizations are
different. This is because the $(u_{\v i\v j},a_0^l\tau^l)$ labeling
of the physical spin wave function $\Phi^{(u_{\v i\v j},a_0^l\tau^l)}$ is not
an one-to-one labeling. Two mean-field ansatz $(u_{\v i\v j},a_0^l\tau^l)$ and
$(\t u_{\v i\v j},\t a_0^l\tau^l)$ differed by an $SU(2)$ gauge
transformation defined by
\begin{align}
\label{GTrans}
 \t u_{\v i\v j} &=  G(\v i) u_{\v i\v j} G^\dag(\v j) ,
&
 \t a^l_0(\v i)\tau^l &=  G(\v i) a^l_0(\v i)\tau^l G^\dag(\v i) ,
\nonumber\\
 \t \psi_{\v i} &=  G(\v i)  \psi_{\v i} ,
&
 G(\v i) & \in SU(2)
\end{align}
give rise to the same spin wave function $ \Phi^{(u_{\v i\v j},a_0^l\tau^l)}
=\Phi^{(\t u_{\v i\v j},\t a_0^l\tau^l)}$.  This is because the up-spin (the
boson), as a bound state of $\psi_1$ and $\psi_2$, is an singlet of the above
gauge $SU(2)$. Thus the spin wave function, as a projected mean-field wave
function, is invariant under the local $SU(2)$ gauge transformation
\Eq{GTrans}.

Due to the many-to-one labeling, an interesting situation appears.  In order
for a spin wave function to have a translation symmetry, its corresponding
ansatz is only required to be translation invariant up to an $SU(2)$ gauge
transformation. That is the ansatz should be invariant under translation
followed by a proper gauge transformation. For example, to have a translation
symmetry in $\v x$ direction, $u_{\v i\v j}$ should satisfy
\begin{align}
\label{GxTx}
 u_{\v i\v j} &=  G_x T_x (u_{\v i\v j}),\ \ \ 
a_0^l(\v i)\tau^l =  G_x T_x (a_0^l(\v i)\tau^l)  ;
\nonumber\\
 T_x (u_{\v i\v j}) &\equiv u_{\v i-\hat{\v x}, \v j -\hat{\v x}},
\ \ \ G_x (u_{\v i\v j}) \equiv G_x(\v i) u_{\v i\v j} G_x^\dag(\v j)  
\end{align}
for a certain $SU(2)$ gauge transformation $G_x(\v i)$.
For two spin wave functions with the same translation symmetry, their ansatz
can be invariant under translation followed by {\em different} gauge
transformations.  Thus two spin liquids with the same symmetry can have
different PSG's.  This indicates that the PSG characterization is more refined
then the SG characterization. PSG can describe those internal
structures that cannot be distinguished by SG. Therefore, we can
use PSG to characterize quantum orders which cannot be completely
characterized by symmetries.

Now let us explicitly write down PSG for some simple ansatz. We will consider
spin liquids with {\em only} translation symmetry. The SG is
generated by two translations in $x$ and $y$ directions
$SG=\{T_x, T_y\}$. The first ansatz is called Z2A ansatz which has a form
$ u_{\v i,\v i+\v m} = u_{\v m}$,
where $u_{\v m} = u^\dag_{-\v m}$ are generic $2\times 2$ complex matrices.
An element in PSG, in general, is formed by the combined transformation as in
\Eq{GxTx}. Including the translation in the $y$-direction, we find the PSG of
the Z2A ansatz is generated by the following transformations $\{G_0, G_xT_x,
G_yT_y\}$, where three gauge transformations $G_0$, $G_x$ and $G_y$ are given
by
$G_0(\v i) = -\tau^0,\ 
G_x(\v i) = \tau^0,\ 
G_y(\v i) = \tau^0$.
Here $\tau^0$ is the $2\times 2$ identity matrix.
Since the ansatz is already invariant under $T_x$ and $T_y$, 
hence $G_x$ and $G_y$ are trivial. 

We would like to point out that a PSG contains a special subgroup, which will
be called the invariant gauge group (IGG). An IGG is formed by pure gauge
transformations that leave the ansatz unchanged
\begin{align}
 IGG\equiv \{ G |\ & 
u_{\v i\v j}= G(\v i) u_{\v i\v j} G^\dag(\v j), \\
& a_0^l(\v i)\tau^l= G(\v i)a_0^l(\v i)\tau^l  G^\dag(\v i)
\} \nonumber 
\end{align}
One can show that PSG, IGG, and SG are related $PSG/IGG=SG$, thus comes the
name projective symmetry group.  For the Z2A ansatz, we find the IGG is a
$Z_2$ group generated by $G_0=-\tau^0$.  Because of this we will call such a
spin liquid a $Z_2$ spin liquid.

Next we consider another ansatz, the Z2B ansatz,
$ u_{\v i,\v i+\v m} = (-1)^{m_x i_y} u_{\v m}$.
The Z2B PSG is still generated by 
$\{G_0, G_xT_x, G_yT_y\}$, but with a different $G_y$:
$G_0(\v i) = -\tau^0,\ 
G_x(\v i) = \tau^0,\ 
G_y(\v i) = (-1)^{i_x} \tau^0$.
Under translation $T_y$,
$
 u_{\v i,\v i+\v m} \to
 u_{\v i-\hat{\v y},\v i+\v m-\hat{\v y}} 
 = (-)^{m_x} u_{\v i,\v i+\v m} .
$
Thus we need a nontrivial gauge transformation $G_y=(-)^{i_x}$ to remove the
extra factor $(-)^{m_x}$. 
The IGG for the Z2B ansatz, generated by $G_0$,
is also $Z_2$ and the ansatz describes another $Z_2$ spin liquid.

One can show that\cite{Wqo} the above Z2A and Z2B PSG's
cannot be transformed into each other by the $SU(2)$ gauge transformation
\Eq{GTrans}. Therefore the Z2A and Z2B ansatz describe two spin
liquids with the same symmetry but different quantum orders.  This
demonstrates that the ground state wave functions of spin liquids contain
structures that cannot be characterized by symmetry. A generalization of the
SG, PSG, can capture some of those extra structures.

To experimentally measure the different quantum orders in the Z2A and Z2B
states, we can measure the spectrum of spin-1 excitations (\ie the two-spinon
excitations). One can show that (see Eq. (96) in \Ref{Wqo}), for the Z2B
state, the spin-1 spectrum is periodic in $\v k$-space with a period $\pi$ in
both $\v x$ and $\v y$ directions. For the Z2A state, the spin-1 spectrum has
the usual period of $2\pi$.

After constructing the spin wave functions of some spin liquids via the
projection \Eq{PhiPsi}, we would like to ask: do those spin liquids actually
exist for spin-1/2 systems? Are there any Hamiltonians such that the
constructed spin wave functions are the ground state of those Hamiltonians?
This is a very hard question and there are some recent work supporting the
existence of spin liquids.\cite{MLB9964,CBP0101,KI0148,CTT0135}
It is also known that spin liquids do exist for certain large $N$
systems.\cite{AM8874,RS9173,Wlight}.
Here we would like to address an easier question about the stability of spin
liquids: can a small perturbation in a spin Hamiltonian change the quantum
order in the corresponding spin liquid. This question can be addressed without
knowing the details of the lattice spin Hamiltonian. We only need to know the
low lying spin excitations and their interactions.

As an example, we consider the stability of the Z2A spin liquid.
Within the $SU(2)$ slave-boson mean-field theory,\cite{AZH8845,DFM8826} the
spin excitations are described by the free spin-1/2 spinons $\psi$ in the
mean-field Hamiltonian \Eq{Hmean}.  In general, if $|\Tr(u_{\v m})|$ is much
less then $|\Tr(u_{\v m}\tau^l)|$, the spinons have finite energy gap. As we
go beyond the mean-field approximation, the low energy excitation also contain
a collective mode described by the fluctuations $\del u_{\v i\v j}$.  The
spinons are no longer free since they interact with $\del u_{\v i\v j}$.  As
pointed out in \Ref{BA8880}, the fluctuations $\del u_{\v i\v j}$ correspond
to a gauge field at low energies.  The gauge group of the low energy gauge
fluctuations is determined by the PSG.\cite{Wqo} In fact it is the IGG of the
ansatz. (See also \Ref{RS9173,Wsrvb,MF9400}.) Thus the $Z_2$ spin liquid has a
$Z_2$ gauge fluctuations at low energies.  The $Z_2$ gauge fluctuations will
cause an interaction between the spinons since they carry unit $Z_2$ charge.
However, the $Z_2$ gauge interaction between the spinons is short ranged. Thus
even beyond the mean-field theory, the spin 1/2 spinon excitations 
are free at low energies and the low energy gauge fluctuations do not cause
any instability. All excitations, spinons and the $Z_2$ gauge fluctuations,
have a finite energy gap. In this way, we showed that the Z2A spin liquid
can be a stable spin liquid which represents a quantum phase.  The
low energy properties and the quantum order of the spin liquid do not change
as we perturb the spin Hamiltonian.

In the above, we only showed that fluctuations around the $Z_2$ mean-field
state do not cause any infrared divergence. However, even short distance
fluctuations can cause instability if they are strong enough. Indeed the short
distance fluctuations in our model are of order $O(1)$. This means that the
projection \Eq{PhiPsi} causes a big change and it is unclear if the projected
state and the original state share similar physical properties. To make our
approach here into a controlled calculation, we need to generalize our model
to some large-$N$ model. One way to do so is to generalize our mean-field
Hamiltonian \Eq{Hmean} to
\begin{equation}
\label{HmeanN}
 H_{mean}= \sum_{\v i\v j} \psi^\dag_{I,\v i} u_{\v i\v j} \psi_{I,\v j}
+ \sum_{\v i} \psi^\dag_{I,\v i} a_0^l(\v i) \tau^l \psi_{I,\v i}
\end{equation}
where $I=1,...,N$ and we have $N$ copies of two-component fields $\psi_I$.
After integrating out the fermions, the effective Lagrangian for the
fluctuations $\del u_{\v i\v j}$ has a form $\cL= N \cL_0(\del u_{\v i\v j})$.
In the large $N$ limit, the fluctuations are suppressed and the mean-field
theory becomes exact.  Thus the two $Z_2$ states discussed above should exist
in those large-$N$ systems.

To obtain the physical system that corresponds to the mean-field theory
\Eq{HmeanN}, we note that a physical state must be a singlet of the gauge
$SU(2)$ on every site. Thus the physical states on each site is formed by the
singlet state of the fermions $\psi_I$. They include $|0\>$,
$\psi^\dag_{\al,I} \psi^\dag_{\bt,J}\eps^{\al\bt}|0\>$, \etc.  The total
number of physical state on each site is $N_p=\sum_{m=0}^{[N/2]} 2^{N-3m}
(2m)!/m!$. Such a system is a system with $N_p$ state per site (which can be
viewed as a spin $S=(N_p-1)/2$ system without any spin rotation
symmetry).

The PSG not only provides a concrete description of quantum order, it also
allows us to partially classify quantum orders.\cite{Wqo} To understand the
significance of PSG, we note that
our understanding of solids is built on two corner stones:
(A) Solids contain an order that is related to broken symmetry.
(B) The order can be described and classified by SG's. In this paper
we propose that a understanding of quantum phase should be built on (at least)
the following two concepts:
(A) A quantum phase in general contain a new kind of order - quantum order -
which may not have broken symmetry and local order parameters. 
(B) The quantum order can be (partially) described and classified 
by PSG's.

%


This research is supported by NSF Grant No. DMR--97--14198.

\bibliography{/home/wen/bib/wencross,/home/wen/bib/htc,/home/wen/bib/misc,/home/wen/bib/fqh,/home/wen/bib/publst}

\begin{thebibliography}{22}
\expandafter\ifx\csname natexlab\endcsname\relax\def\natexlab#1{#1}\fi
\expandafter\ifx\csname bibnamefont\endcsname\relax
  \def\bibnamefont#1{#1}\fi
\expandafter\ifx\csname bibfnamefont\endcsname\relax
  \def\bibfnamefont#1{#1}\fi
\expandafter\ifx\csname citenamefont\endcsname\relax
  \def\citenamefont#1{#1}\fi
\expandafter\ifx\csname url\endcsname\relax
  \def\url#1{\texttt{#1}}\fi
\expandafter\ifx\csname urlprefix\endcsname\relax\def\urlprefix{URL }\fi
\providecommand{\bibinfo}[2]{#2}
\providecommand{\eprint}[2][]{\url{#2}}

\bibitem[{\citenamefont{Landau and Lifschitz}(1958)}]{LL58}
\bibinfo{author}{\bibfnamefont{L.~D.} \bibnamefont{Landau}} \bibnamefont{and}
  \bibinfo{author}{\bibfnamefont{E.~M.} \bibnamefont{Lifschitz}},
  \emph{\bibinfo{title}{Satistical Physics - Course of Theoretical Physics Vol
  5}} (\bibinfo{publisher}{Pergamon}, \bibinfo{address}{London},
  \bibinfo{year}{1958}).

\bibitem[{\citenamefont{Ginzburg and Landau}(1950)}]{GL5064}
\bibinfo{author}{\bibfnamefont{V.~L.} \bibnamefont{Ginzburg}} \bibnamefont{and}
  \bibinfo{author}{\bibfnamefont{L.~D.} \bibnamefont{Landau}},
  \bibinfo{journal}{J. exp. theor. Phys.} \textbf{\bibinfo{volume}{20}},
  \bibinfo{pages}{1064} (\bibinfo{year}{1950}).

\bibitem[{\citenamefont{Wen}(1989)}]{Wtop}
\bibinfo{author}{\bibfnamefont{X.-G.} \bibnamefont{Wen}},
  \bibinfo{journal}{Phys. Rev. B} \textbf{\bibinfo{volume}{40}},
  \bibinfo{pages}{7387} (\bibinfo{year}{1989}).

\bibitem[{\citenamefont{Wen and Niu}(1990)}]{WNtop}
\bibinfo{author}{\bibfnamefont{X.-G.} \bibnamefont{Wen}} \bibnamefont{and}
  \bibinfo{author}{\bibfnamefont{Q.}~\bibnamefont{Niu}},
  \bibinfo{journal}{Phys. Rev. B} \textbf{\bibinfo{volume}{41}},
  \bibinfo{pages}{9377} (\bibinfo{year}{1990}).

\bibitem[{\citenamefont{Wen}(1995)}]{Wtoprev}
\bibinfo{author}{\bibfnamefont{X.-G.} \bibnamefont{Wen}},
  \bibinfo{journal}{Advances in Physics} \textbf{\bibinfo{volume}{44}},
  \bibinfo{pages}{405} (\bibinfo{year}{1995}).

\bibitem[{\citenamefont{Wen}(2001)}]{Wqo}
\bibinfo{author}{\bibfnamefont{X.-G.} \bibnamefont{Wen}},
  \bibinfo{journal}{cond-mat/0107071, to appear in PRB}
  (\bibinfo{year}{2001}).

\bibitem[{\citenamefont{Lifshitz}(1960)}]{L6030}
\bibinfo{author}{\bibfnamefont{I.~M.} \bibnamefont{Lifshitz}},
  \bibinfo{journal}{Sov. Phys. JETP} \textbf{\bibinfo{volume}{11}},
  \bibinfo{pages}{1130} (\bibinfo{year}{1960}).

\bibitem[{\citenamefont{Affleck et~al.}(1988)\citenamefont{Affleck, Zou, Hsu,
  and Anderson}}]{AZH8845}
\bibinfo{author}{\bibfnamefont{I.}~\bibnamefont{Affleck}},
  \bibinfo{author}{\bibfnamefont{Z.}~\bibnamefont{Zou}},
  \bibinfo{author}{\bibfnamefont{T.}~\bibnamefont{Hsu}}, \bibnamefont{and}
  \bibinfo{author}{\bibfnamefont{P.~W.} \bibnamefont{Anderson}},
  \bibinfo{journal}{Phys. Rev. B} \textbf{\bibinfo{volume}{38}},
  \bibinfo{pages}{745} (\bibinfo{year}{1988}).

\bibitem[{\citenamefont{Dagotto et~al.}(1988)\citenamefont{Dagotto, Fradkin,
  and Moreo}}]{DFM8826}
\bibinfo{author}{\bibfnamefont{E.}~\bibnamefont{Dagotto}},
  \bibinfo{author}{\bibfnamefont{E.}~\bibnamefont{Fradkin}}, \bibnamefont{and}
  \bibinfo{author}{\bibfnamefont{A.}~\bibnamefont{Moreo}},
  \bibinfo{journal}{Phys. Rev. B} \textbf{\bibinfo{volume}{38}},
  \bibinfo{pages}{2926} (\bibinfo{year}{1988}).

\bibitem[{\citenamefont{Jain}(1991)}]{J9153}
\bibinfo{author}{\bibfnamefont{J.~K.} \bibnamefont{Jain}},
  \bibinfo{journal}{Phys. Rev. B} \textbf{\bibinfo{volume}{41}},
  \bibinfo{pages}{7653} (\bibinfo{year}{1991}).

\bibitem[{\citenamefont{Wen}(1999)}]{Wpc}
\bibinfo{author}{\bibfnamefont{X.-G.} \bibnamefont{Wen}},
  \bibinfo{journal}{Phys. Rev. B} \textbf{\bibinfo{volume}{60}},
  \bibinfo{pages}{8827} (\bibinfo{year}{1999}).

\bibitem[{\citenamefont{Gutzwiller}(1963)}]{G6359}
\bibinfo{author}{\bibfnamefont{M.~C.} \bibnamefont{Gutzwiller}},
  \bibinfo{journal}{Phys. Rev. Lett.} \textbf{\bibinfo{volume}{10}},
  \bibinfo{pages}{159} (\bibinfo{year}{1963}).

\bibitem[{\citenamefont{Misguich et~al.}(1999)\citenamefont{Misguich,
  Lhuillier, Bernu, and Waldtmann}}]{MLB9964}
\bibinfo{author}{\bibfnamefont{G.}~\bibnamefont{Misguich}},
  \bibinfo{author}{\bibfnamefont{C.}~\bibnamefont{Lhuillier}},
  \bibinfo{author}{\bibfnamefont{B.}~\bibnamefont{Bernu}}, \bibnamefont{and}
  \bibinfo{author}{\bibfnamefont{C.}~\bibnamefont{Waldtmann}},
  \bibinfo{journal}{Phys. Rev. B} \textbf{\bibinfo{volume}{60}},
  \bibinfo{pages}{1064} (\bibinfo{year}{1999}).

\bibitem[{\citenamefont{Kashima and Imada}(2001)}]{KI0148}
\bibinfo{author}{\bibfnamefont{T.}~\bibnamefont{Kashima}} \bibnamefont{and}
  \bibinfo{author}{\bibfnamefont{M.}~\bibnamefont{Imada}},
  \bibinfo{journal}{cond-mat/0104348}
  (\bibinfo{year}{2001}).

\bibitem[{\citenamefont{Coldea et~al.}(2001)\citenamefont{Coldea, Tennant,
  Tsvelik, and Tylczynski}}]{CTT0135}
\bibinfo{author}{\bibfnamefont{R.}~\bibnamefont{Coldea}},
  \bibinfo{author}{\bibfnamefont{D.}~\bibnamefont{Tennant}},
  \bibinfo{author}{\bibfnamefont{A.}~\bibnamefont{Tsvelik}}, \bibnamefont{and}
  \bibinfo{author}{\bibfnamefont{Z.}~\bibnamefont{Tylczynski}},
  \bibinfo{journal}{Phys. Rev. Lett.} \textbf{\bibinfo{volume}{86}},
  \bibinfo{pages}{1335} (\bibinfo{year}{2001}).

\bibitem[{\citenamefont{Capriotti et~al.}(2001)\citenamefont{Capriotti, Becca,
  Parola, and Sorella}}]{CBP0101}
\bibinfo{author}{\bibfnamefont{L.}~\bibnamefont{Capriotti}},
  \bibinfo{author}{\bibfnamefont{F.}~\bibnamefont{Becca}},
  \bibinfo{author}{\bibfnamefont{A.}~\bibnamefont{Parola}}, \bibnamefont{and}
  \bibinfo{author}{\bibfnamefont{S.}~\bibnamefont{Sorella}},
  \bibinfo{journal}{Phys. Rev. Lett.} \textbf{\bibinfo{volume}{87}},
  \bibinfo{pages}{097201} (\bibinfo{year}{2001}).

\bibitem[{\citenamefont{Read and Sachdev}(1991)}]{RS9173}
\bibinfo{author}{\bibfnamefont{N.}~\bibnamefont{Read}} \bibnamefont{and}
  \bibinfo{author}{\bibfnamefont{S.}~\bibnamefont{Sachdev}},
  \bibinfo{journal}{Phys. Rev. Lett.} \textbf{\bibinfo{volume}{66}},
  \bibinfo{pages}{1773} (\bibinfo{year}{1991}).

\bibitem[{\citenamefont{Affleck and Marston}(1988)}]{AM8874}
\bibinfo{author}{\bibfnamefont{I.}~\bibnamefont{Affleck}} \bibnamefont{and}
  \bibinfo{author}{\bibfnamefont{J.~B.} \bibnamefont{Marston}},
  \bibinfo{journal}{Phys. Rev. B} \textbf{\bibinfo{volume}{37}},
  \bibinfo{pages}{3774} (\bibinfo{year}{1988}).

\bibitem[{\citenamefont{Wen}(2002)}]{Wlight}
\bibinfo{author}{\bibfnamefont{X.-G.} \bibnamefont{Wen}},
  \bibinfo{journal}{Phys. Rev. Lett.} \textbf{\bibinfo{volume}{88}},
  \bibinfo{pages}{11602} (\bibinfo{year}{2002}).

\bibitem[{\citenamefont{Baskaran and Anderson}(1988)}]{BA8880}
\bibinfo{author}{\bibfnamefont{G.}~\bibnamefont{Baskaran}} \bibnamefont{and}
  \bibinfo{author}{\bibfnamefont{P.~W.} \bibnamefont{Anderson}},
  \bibinfo{journal}{Phys. Rev. B} \textbf{\bibinfo{volume}{37}},
  \bibinfo{pages}{580} (\bibinfo{year}{1988}).

\bibitem[{\citenamefont{Wen}(1991)}]{Wsrvb}
\bibinfo{author}{\bibfnamefont{X.-G.} \bibnamefont{Wen}},
  \bibinfo{journal}{Phys. Rev. B} \textbf{\bibinfo{volume}{44}},
  \bibinfo{pages}{2664} (\bibinfo{year}{1991}).

\bibitem[{\citenamefont{Mudry and Fradkin}(1994)}]{MF9400}
\bibinfo{author}{\bibfnamefont{C.}~\bibnamefont{Mudry}} \bibnamefont{and}
  \bibinfo{author}{\bibfnamefont{E.}~\bibnamefont{Fradkin}},
  \bibinfo{journal}{Phys. Rev. B} \textbf{\bibinfo{volume}{49}},
  \bibinfo{pages}{5200} (\bibinfo{year}{1994}).

\end{thebibliography}

\end{document}